\newcommand{\LASCUO}{{$\mathrm{La_{1.85}Sr_{0.15}CuO_{4+\delta}}$ }} 
\newcommand{\LACUOD}{{$\mathrm{La_{2}CuO_{4+\delta}}$ }} 
\newcommand{\YBCO}{{$\mathrm{YBa_{2}Cu_{3}O_{6+\delta}}$ }} 
\newcommand{\NCCO}{{$\mathrm{Nd_{1.86}Ce_{0.14}CuO_{4+\delta}}$ }} 
 
\newcommand{\NCO}{{$\mathrm{Nd_2CuO_{4+\delta}}$ }} 
 
\newcommand{\ixs}{inelastic x-ray scattering } 

\documentclass[aps,prl,twocolumn,superscriptaddress]{revtex4} 
\usepackage{graphicx}  
\usepackage{dcolumn}% Align table columns on decimal point
\usepackage{bm}% bold math
\begin{document}  
  
%\preprint{}  
  
\title{Anomalous Dispersion of Longitudinal Optical Phonons in  
$\mathrm{\mathbf{Nd_{1.86}Ce_{0.14}CuO_{4+\bm{\delta}}}}$
Determined by Inelastic X-ray Scattering}  
  
  
\author{M. d'Astuto}  
%\email[]{astuto@esrf.fr}  
\affiliation{European Synchrotron Radiation Facility, 
BP 220, F-38043 Grenoble Cedex, France}  
%\thanks{}  
\author{P. K. Mang}  
\affiliation{Department of Applied
Physics, Stanford University, Stanford, California 94305}  
\author{P. Giura}  
\author{A. Shukla} 
\affiliation{European Synchrotron Radiation Facility, 
BP 220, F-38043 Grenoble Cedex, France}  
\author{P. Ghigna} 
\affiliation{Dipartimento di Chimica Fisica ``M. Rolla'', 
Un. Pavia, V.le Taramelli 16, I-27100, Pavia, Italy}  
\author{A. Mirone}  
\affiliation{European Synchrotron Radiation Facility, 
BP 220, F-38043 Grenoble Cedex, France}  
\author{M. Braden}  
\affiliation{II. Physikalisches Inst., 
Univ. zu K\"{o}ln, Z\"{u}lpicher Str. 77, 50397 K\"{o}ln, Germany}  
\author{M. Greven}
\affiliation{Department of Applied
Physics, Stanford University, Stanford, California 94305}    
\affiliation{Stanford Synchrotron Radiation 
Laboratory, Stanford, California 94305}
\author{M. Krisch} 
\author{F. Sette} 
\affiliation{European Synchrotron Radiation Facility, 
BP 220, F-38043 Grenoble Cedex, France}  
  
\date{\today}  
  
\begin{abstract}  
The phonon dispersions of \NCCO
along the $[\xi,0,0]$   
direction have been determined by inelastic x-ray scattering.  
Compared to the undoped parent compound, the two highest longitudinal 
phonon branches, associated with the Cu-O bond-stretching 
and out-of-plane oxygen vibration, are shifted to lower energies. 
Moreover, an anomalous softening of the bond-stretching 
band is observed around $\mathbf{q}=(0.2,0,0)$. These signatures provide  
evidence for strong electron-phonon coupling in 
this electron-doped high-temperature superconductor. 

\end{abstract}  
  
%\keywords{}  
  
\maketitle  
  
% References should be done using the  \cite, \ref, and \label commands  
%\section{}  
While the coupling between electrons and phonons is known to be 
the driving mechanism 
for Cooper-pair formation in conventional superconductors, its 
role in the high-critical-temperature 
superconductors (HTcS) is the subject of intense research 
efforts.
Recently, evidence for electron-phonon coupling
has been invoked in the interpretation of 
inelastic neutron scattering 
(INS) and angle-resolved photoemission spectroscopy (ARPES) experiments.
The INS studies, carried out 
on \LASCUO \cite{pintscB,pingin,mcqueeny,pinbrief},
oxygen-doped \LACUOD \cite{lacuod} and 
\YBCO  \cite{pingin,reichardt} reveal an 
anomalous softening with doping of the highest longitudinal 
optical (LO) phonon branch,
in particular along the $q=[\xi,0,0]$ 
direction. 
This branch is 
 assigned to the Cu-O bond-stretching mode \cite{raman,pingin}.  
The observed softening has been interpreted as a signature of a 
strong electron-phonon coupling \cite{pingin,mcqueeny}, which has been 
discussed since the discovery of HTcS \cite{bednorz,weber}. 
Furthermore, in an energy range similar to the LO bond-stretching phonon 
band, ARPES studies
on three different families of hole-doped 
HTcS reveal a distinct ``kink'' anomaly in the 
quasiparticle dispersion  \cite{lanzara}. 
The scenario emerging from the above INS and ARPES works suggests a 
strong coupling between the charge carriers and the Cu-O bond-stretching 
phonon modes to be ubiquitous in HTcS materials, at least for hole-doped
compounds. However,  
its role in the pairing mechanism remains completely
unclear \cite{pinbrief}. At the moment, it may not even be excluded that
the electron-phonon interaction is pair-breaking for 
the d-wave superconducting order parameter.
Therefore, it appears very important to analyze
the strength of the phonon anomalies in as many 
cuprate families as possible and to compare them with their
superconducting properties.
In this context, the electron-doped cuprates are of central importance
due to the distinct character of the doped 
charges in this material:  Cu $3d_{x^2-y^2}$ (O $2p$) for n(p)-type
cuprates \cite{ttu}, leading to a
very different electronic structure \cite{armitage}.
Since the phonon anomalies are related to a coupling 
between charge fluctuations and phonons, charges with different
character may induce quite different electron-phonon interactions.

In this Letter, we present an \ixs (IXS) study of the
phonon dispersion in the n-type
cuprates \NCCO (NCCO). 
Inelastic x-ray scattering can overcome the main limitation of
inelastic neutron scattering, {\it i.e.} the need for sufficiently 
large single crystals of high chemical 
and structural quality \cite{pingin}.
Lateral x-ray beam sizes of few tens of $\mu$m are routinely obtained. 
Moreover, at photon 
energies around 10-20 keV and $Z >$ 3, 
the total cross section is dominated 
by photoelectric absorption, and therefore the  
typical x-ray penetration depths for high-$Z$ materials is of the order of 
10 - 100 $\mu$m.
Consequently, very small samples (down to less than $10^{-4}$ mm$^{3}$) 
can be studied with signal rates comparable to typical INS 
experiments on cm$^3$-sized samples.
Despite these advantages, little work has been done using IXS 
on the HTcS compounds \cite{burkel}.
We choose $\mathrm{Nd_{2-x}Ce_{x}CuO_{4+\delta}}$  
for our IXS study, since its crystallographic 
structure is one of the simplest among the HTcS, and because  
extensive INS studies exist for its undoped parent compound 
\NCO (NCO) \cite{pintscB,pingin}.  
Our interest is focused on the 
$[\xi,0,0]$ direction, where the LO branch displays its strongest anomaly 
for hole-doped HTcS \cite{mcqueeny,pinbrief}. 
The present results reveal that, near the zone center, the two 
highest longitudinal optical branches, assigned to the Cu-O  
bond-stretching and O(2) vibration modes, are shifted to 
lower frequencies with respect to the undoped parent compound. 
The interpretation of our data 
is supported by lattice dynamics calculations, taking 
into account a \textit{Thomas-Fermi} screening mechanism. 
Furthermore, we observe an anomalous softening of the highest branch 
around $\mathbf{q}=(0.2,0,0)$. Our results demonstrate that 
this anomalous behavior of the high-energy LO phonon branch 
is a universal property of both hole- and 
electron-doped HTcS compounds, therefore providing further evidence 
that electron-phonon interactions may play an important role in
high-Tc superconductivity. Furthermore, the present results are an 
important demonstration of IXS as a powerful tool for the 
study of the lattice dynamics in small, high-quality 
crystals of complex transition metal oxides.

The experiment was carried out at the very-high-energy-resolution IXS 
beam-line ID16 at the European Synchrotron Radiation Facility (ESRF).  
X-rays from an undulator source were monochromated using a Si (111) 
double-crystal monochromator, followed by 
a high-energy-resolution backscattering 
monochromator \cite{ixsmono}, operating at 15816 eV 
(Si (888) reflection order). A toroidal gold-coated mirror 
refocused the x-ray beam onto the sample, where a beam size of 
$250 \times 250~ \mu$m$^2$ full-width-half-maximum (FWHM) was obtained. 
The scattered photons were energy-analyzed by a 
spherical silicon crystal analyzer 3 m in radius, 
operating at the same Bragg reflection  
as the monochromator \cite{ixsana}. The total energy resolution 
was 1.6 THz (6.6 meV) FWHM. The momentum transfer $\mathbf{Q}$ 
was selected by rotating the 3 m spectrometer arm in the 
scattering plane perpendicular to the linear x-ray polarization vector 
of the incident beam. 
The momentum resolution was set to 
$\approx 0.087~\mathrm{\AA^{-1}}$ in both 
the horizontal and the vertical direction by an aperture of 
$20\times20$ mm$^2$ in front of the analyzer. 
Further experimental details are given elsewhere (\cite{ixsmono,ixsana} 
and references therein).
The sample is a single crystal grown by the traveling-solvent floating-zone
method in 4 atm of $\mathrm{O_2}$ at Stanford University. 
It has been reduced under pure Ar atmosphere at $920^{\circ} $C for 20 h, 
followed by a further 20 h of exposure at $500^{\circ}$C to a pure 
$\mathrm{O_2}$ atmosphere. 
Such a procedure is necessary to produce 
a superconducting phase in $\mathrm{Nd_{1.86}Ce_{0.14}CuO_{4+\delta}}$, 
although its exact effect is not understood. 
Following this treatment the sample had 
a narrow superconducting transition with an onset 
temperature of $T_c=24$ K.
The sample is of very good crystalline quality,  
with a rocking curve width of 0.02$^{\circ}$ (FWHM) 
around the $[h,0,0]$ direction. 
It was mounted on 
the cold finger of a closed-loop helium cryostat, and cooled to 
15 K. The experiment was performed in reflection geometry, and the 
probed scattering volume corresponded to about $1.5\times10^{-3}$ mm$^3$.  
IXS scans were performed in the -2$<\nu<$24 THz range,  
in the $\bm{\tau}=(6,0,0)$ and $\bm{\tau}=(7,1,0)$ Brillouin zones  
\footnote{All reciprocal lattice vectors are expressed in 
r.l.u.}. 
The data were collected along the following three lines: \\
I) $\mathbf{Q}=(6+\xi,0,0)$, in longitudinal configuration 
({\textit i.e.} with $\mathbf{q}=(\xi,0,0)~ \parallel\mathbf{Q}$), \\
II) $\mathbf{Q}=(7\pm\xi,1,0)$  in almost longitudinal 
configuration ({\textit i.e.} with 
$\mathbf{q} = (\xi,0,0) $ and $(\mathbf{Q} \cdot \mathbf{q})/Q \approx q$),\\ 
III) $\mathbf{Q}=(7,1-\xi,0)$ in  
almost transverse configuration 
($\mathbf{q} = (0,\xi,0)$ and $(\mathbf{Q} \cdot \mathbf{q})/Q \approx 0$).\\ 
The low temperature and high momentum transfer were chosen so as to 
optimize the count rate on the high-frequency optical mode while 
limiting the loss of contrast due to the contribution from the tails of 
the intense low frequency acoustic modes.

\begin{figure}  
\includegraphics[angle=-90,scale=0.30]{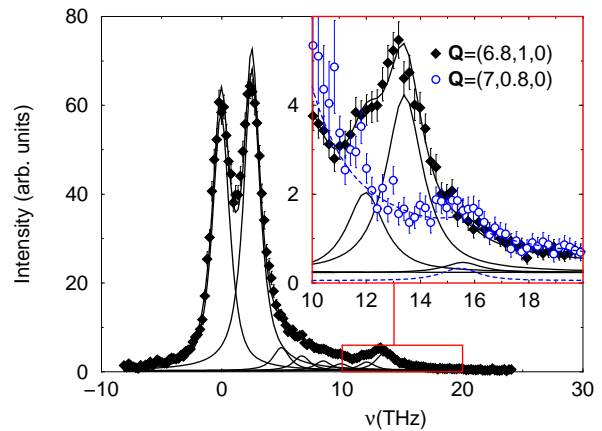}  
\caption{\label{figa} Experimental IXS phonon   
spectra of \NCCO at $T=15$ K and corresponding harmonic 
oscillator model best fits (solid and dotted lines). 
Both scans were performed in the  
$\bm{\tau}=(7, 1, 0)$ Brillouin zones   
with a propagation vector of $\xi=0.2$ in an almost longitudinal 
geometry along the $a^*$ direction (diamonds) and in an almost 
transverse geometry along $b^*$ (open circles).} 
\end{figure}  

Fig. \ref{figa} shows a typical energy scan in almost 
longitudinal geometry at $\mathbf{Q}=(6.8,1,0)$, corresponding 
to $\xi=0.2$. 
The data are shown together with the results of a fit, 
where the excitations where modeled by harmonic oscillators, convoluted 
with the instrumental resolution function. 
Three features can be clearly distinguished: 
(i) an elastic peak at 0 THz, due to  
chemical disorder and, possibly, strain due to the different 
thermal expansion  
coefficients between the sample and the sample support during cooling, 
(ii) the longitudinal acoustic phonon, centered near 2.7 THz and (iii) a 
weaker feature around  13~THz. 
In the intermediate energy region (from about 4 to 10 THz), no distinct 
phonon peaks are resolved due to the 
dominating contribution from the tails of the elastic and acoustic 
phonon signals. 
The inset of Fig. 1 emphasizes the high-energy region 
of the spectrum. One can clearly 
distinguish two phonons, centered around 12 and 13.5 THz, respectively. 
The weak shoulder at around 16 THz can be attributed to an
admixture of the transverse optical (TO) mode, as indicated by the comparison
(after normalization to the same intensity at 20 THz) with the equivalent
scan in transverse geometry $\mathbf{Q}=(7,0.8,0)$. 
Consequently, the 
two stronger peaks are unambiguously assigned to longitudinal optical 
(LO) modes. From the absence of other higher frequency modes up to 23 THz, 
we conclude that the two modes at 12 and 13.5 THz can be 
identified with the O(2) vibration and Cu-O bond-stretching 
modes, respectively (see Refs. \onlinecite{raman,pingin}). 

\begin{figure}  
\includegraphics[scale=0.30]{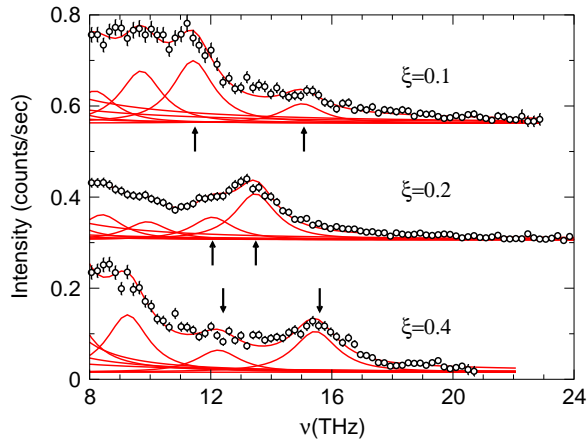}  
\caption{\label{figb} IXS spectra 
in the $\bm{\tau}=(7, 1, 0)$ Brillouin zone   
with propagation vector $\mathbf{q}$ along $a^*$, as indicated in the figure. 
The experimental data (circles) are shown together with their 
corresponding harmonic oscillator model  
best fits (solid lines), as discussed in the text.}
\end{figure}  

In order to determine the dispersion of the two highest LO branches, 
IXS spectra were recorded for $0<\xi\le1$.
In Fig. \ref{figb} we show three spectra taken along the 
$(7-\xi,1,0)$ direction. 
Close to the zone center, at $\mathbf{q}=(0.1, 0, 0 )$, the highest 
frequency mode is observed slightly above 15 THz. 
At $\mathbf{q}=(0.2, 0, 0 )$ the highest mode is found at the much 
lower energy of 13.5 THz. Finally, for $\mathbf{q}=(0.4, 0, 0)$,
we again find a high-frequency mode around 15.5 THz. 

\begin{figure}  
\includegraphics[scale=0.35]{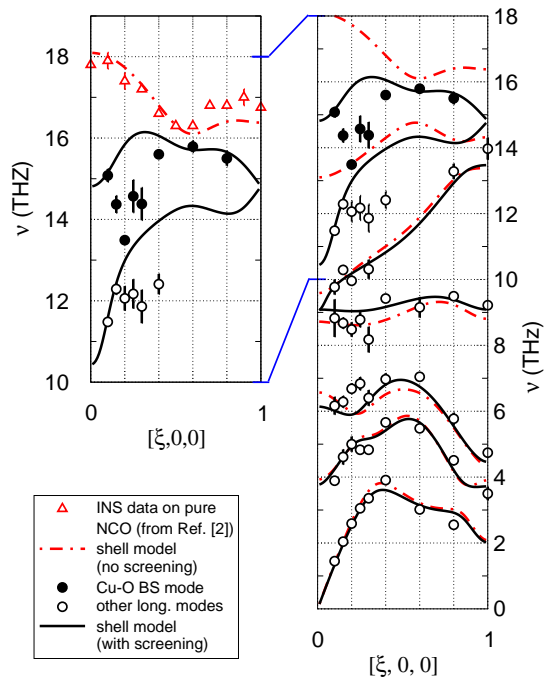}  
\caption{\label{figc} Right hand side: 
($\circ$, $\bullet$) experimental longitudinal 
phonon frequencies determined from IXS spectra in  
NCCO at $T=15$ K along the $[\xi,0,0]$  direction.
Solid circles ($\bullet$) emphasize the highest energy frequencies
measured.
Solid (dot-dashed) lines indicate lattice dynamics calculation 
with a screened (unscreened) Coulomb interaction. \\
Left hand side: magnification of the high energy portion of the right hand 
side graph showing the dispersion of the top two phonon branches.
The highest frequency branch ($\triangle$), as measured by INS (from Ref. 
\onlinecite{pintscB}) in the insulating parent compound \NCO is shown 
for comparison.  
The experimental data are in agreement with calculations, 
except for the anomalous softening at $q=(0.2,0,0)$ of the
two higher energy branches (see text).} 
\end{figure}  

The peak positions extracted from these and many other scans
are summarized in Fig. \ref{figc}. The highest branch exhibits a sharp
dip around $\mathbf{q}=(0.2, 0, 0)$ and recovers for larger $q$-values. 
This behavior is most likely due to an anti-crossing with the 
second highest branch which is mainly
associated with vibrations of the O(2) position in the $\xi$-direction. 
Therefore, within a standard anti-crossing framework, one should 
interpret the highest longitudinal intensities 
for $\mathbf{q}=(0.3, 0, 0)$ and above as being mainly
due to O(2) vibration. 
Within that scenario, this second-highest branch 
increases its frequency in the middle of the zone 
as in the undoped compound, and, except 
for the fact that the \textit{Lyddane-Sachs-Teller} (LO-TO) gap closes,  
seems to be insensitive to doping. 
The LO bond-stretching mode just above $\mathbf{q}=(0.2, 0, 0)$ 
is then found at quite low energies, $\sim$ 12 THz, but can not be 
unambiguously followed to larger q-values.
Nevertheless, our data document that the LO bond-stretching branch in NCCO
is strongly renormalized compared to undoped NCO, in particular it
bends down anomalously from the zone center 
to $\mathbf{q}=(0.2, 0, 0)$.

In order to further validate the correctness of our assignments, 
we performed a lattice dynamical calculation \cite{mirone} based on 
a shell model. We used a common potential model for cuprates, 
in which the interatomic potentials have been derived from a comparison  
of INS results for different   
HTcS compounds by Chaplot {\it et al.} \cite{chaplot}, using 
a screened Coulomb potential, 
in order to simulate the effect of the free carriers introduced by doping. 
Following Ref. \cite{chaplot}   
for metallic \LASCUO and $\mathrm{YBa_2Cu_3O_7}$, we replaced the 
long-range Coulomb potential $V_c(q)$ by $V_c(q)/\epsilon(q)$, and for 
the dielectric function we take the semi-classical 
\textit{Thomas-Fermi} limit of $\epsilon(q)=1+\kappa_s^2/q^2$,
where $\kappa_s^2$ indicates the screening vector. 
The results of the calculation (without screening: dot-dashed lines; 
with screening: solid lines) are shown as well in Fig. \ref{figc}. 
The lattice dynamics calculations without 
screening have been included, since they reproduce very well the 
experimental dispersion of the undoped parent compound 
\cite{pintscB}.  
The shift at the zone center of the high-energy phonon branches 
of NCCO  with respect to NCO
is due to the closing of a large LO-TO  
splitting. 
Indeed, the corresponding 
$\Delta_1$ and $\Delta_3$ branches in \NCO 
are separated by almost 3 THz at the zone center, 
as observed by INS \cite{pintscB}. 
We point out that in our case a strong softening due to  
\textit{Thomas-Fermi} screening does not imply a higher 
\textit{Thomas-Fermi}  
parameter $\kappa_s$: actually, we find a screening vector $\kappa_s$ 
of about $0.39~\mathrm{\AA}^{-1}$, which is
comparable to that for \LASCUO \cite{chaplot}.  
Though our modified calculation reproduces the closing of the LO-TO
splitting, we still observe an anomalous additional softening of the
highest bond-stretching LO branch near $\mathbf{q}=(0.2,0,0)$,
which is not reproduced by our calculations (see Fig. \ref{figc}).
This branch softens  
in frequency from  $\mathbf{q}=(0.1, 0, 0)$  to $\mathbf{q}=(0.2, 0, 0)$ 
by about $\Delta\nu\approx$ 1.5 THz, which is a 
shift comparable to the anomalous shift observed in \LASCUO at slightly 
larger $\mathbf{q}$ \cite{pintscB,mcqueeny,pinbrief}.  
Therefore, we believe that this anomalous softening is of the same nature  
as the one observed in p-type $\mathrm{La_{1.85}Sr_{0.15}CuO_{4+\delta}}$.

\begin{figure}  
\includegraphics[scale=0.30]{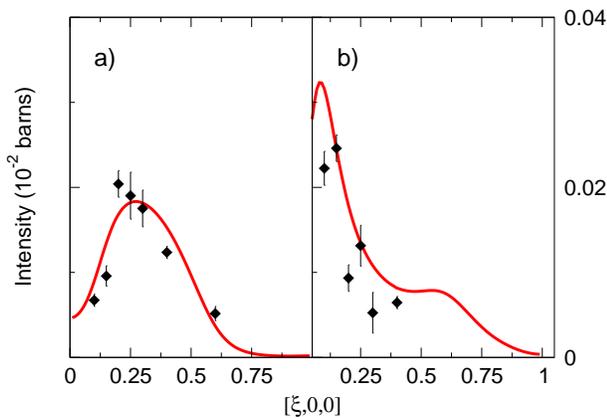}  
\caption{\label{figd}  
Comparison between experimental (diamonds) and calculated (line) 
phonon intensities for the two highest optical phonon branches 
along $(7-\xi,1,0)$. a) bond-stretching LO mode starting around 15 THz and
b) O(2) vibration LO branch starting around 10.5 THz.}
\end{figure}

A comparison between the experimental and calculated intensities 
for the two highest phonon branches is shown in Fig. \ref{figd}. 
The good agreement of the observed integrated intensities with 
the calculated ones for the upper branches 
validates the correctness of our phonon branch assignment, at least for
$\xi \le 0.2$. For $\xi > 0.2$ we would have expected an intensity 
exchange between the two highest branches, which seems to be not observed.

The main difference between \LASCUO and \NCCO is, 
besides the screening effect, that in NCCO the Cu-O bond-stretching branch 
is closer in energy to the out-of-plane oxygen vibration one, 
having almost the same energy at $\xi = 0.2$.
These two branches belong to the same symmetry and therefore cannot cross,
so that for $\xi>0.2$ softening implies interaction with the 
out-of-plane oxygen vibration mode with the same symmetry. 
In the region between $\mathbf{q}=(0.25,0,0)$ and $(0.3,0,0)$, 
the two modes are poorly defined
in energy, which is consistent with 
what is observed in \LASCUO 
by Pintschovius and Braden \cite{pinbrief} 
and McQueeney \textit{et al.} \cite{mcqueeny} for
$\xi\sim 0.25-0.3$. 
The corresponding real space periodicity of 
3 to 4 unit cells may therefore be linked to the proximity
to some charge instability. 
We remark that a reduced vector $\xi\sim 0.25-0.3$ approximately
corresponds to the nesting vector along $[\xi 0 0]$ direction, as can be
inferred from the ARPES data of Ref. \onlinecite{armitage}. 

In conclusion, the present results  
reveal that the anomalous softening previously observed  
in hole-doped compounds 
\cite{pintscB,mcqueeny,pinbrief,lacuod,pingin,reichardt}, 
is also present in the electron-doped cuprates. This is evidenced 
by the comparison of
the present results on doped NCCO with the previously reported 
ones on pure NCO \cite{pintscB}. 
This implies that:
(i) the anomaly also exists in n-type cuprates, 
giving strength to the 
hypothesis \cite{pintscB,mcqueeny,pinbrief,lacuod,pingin,reichardt}
of an electron-phonon coupling origin of this feature; 
(ii) this is a generic feature of the high-temperature superconductors, 
as expected, if phonons are relevant
to high temperature superconductivity. 
 Moreover, this Letter demonstrates that high-energy resolution
inelastic x-ray scattering has developed into an invaluable tool for the
study of the lattice dynamics of complex transition metal oxides, allowing
measurements on small high-quality single crystals which are inaccessible
to the traditional method of inelastic neutron scattering.

%\section{Experiments}  
%\subsection{Samples}  
%\subsection{IXS}  
%\subsubsection{} 
% If in two-column mode, this environment will change to single-column  
% format so that long equations can be displayed. Use  
%\begin{widetext}  
%\end{widetext}  
  
% figures should be put into the text as floats.  
% Use the graphics or graphicx packages (distributed with LaTeX2e)  
% and the \includegraphics macro defined in those packages.  
% See the LaTeX Graphics Companion by Michel Goosens, Sebastian Rahtz,  
% that you will use with \ref{} command in the braces 
%of the \label{} command.  
% entire page. There is no need to do explicit centering.  
  
% \begin{figure}  
% \end{figure}  
  
% Surround figure environment with turnpage environment for landscape  
% figure  
% \begin{figure}  
% \end{figure}  
  
\begin{acknowledgments}  
We acknowledge L. Paolasini and G. Monaco for useful discussions 
and H. Casalta for precious help during preliminary tests. 
The authors are grateful to D. Gambetti, C. Henriquet and R. Verbeni 
for technical help, to J.-L. Hodeau for  
help in the crystal orientation 
and J. -P. Vassalli for crystal cutting.  
P.K.M. and M.G. are supported 
by the U.S. Department of Energy under
Contracts No. DE-FG03-99ER45773 and No. DE-AC03-76SF00515, by NSF
CAREER Award No. DMR-9985067, and by the A.P. Sloan Foundation.  
\end{acknowledgments}

\end{document}